\documentclass[prl,showpacs,showkeys,byrevtex,twocolumn]{revtex4}%
\usepackage{amsfonts}
\usepackage{amsmath}
\usepackage{amssymb}
\usepackage{graphicx}%
\providecommand{\U}[1]{\protect\rule{.1in}{.1in}}
\newtheorem{theorem}{Theorem}

\newtheorem{conjecture}{Conjecture}
\newtheorem{corollary}{Corollary}

\newtheorem{example}{Example}

\newtheorem{remark}{Remark}

\newenvironment{proof}[1][Proof]{\noindent\textbf{#1.} }{\ \rule{0.5em}{0.5em}}
\begin{document}
\preprint{ }
\title[ ]{Optimal Entanglement Formulas for Entanglement-Assisted Quantum Coding}
\author{Mark M. Wilde}
\email{mark.wilde@usc.edu}
\author{Todd A. Brun}
\affiliation{Center for Quantum Information Science and Technology, Communication Sciences
Institute, Department of Electrical Engineering, University of Southern
California, Los Angeles, California 90089, USA}

\begin{abstract}
We provide several formulas that determine the optimal number of entangled
bits\ (ebits) that a general entanglement-assisted quantum code requires. Our
first theorem gives a formula that applies to an arbitrary
entanglement-assisted block code. Corollaries of this theorem give formulas
that apply to a code imported from two classical binary block codes, to a code
imported from a classical quaternary block code, and to a continuous-variable
entanglement-assisted quantum block code. Finally, we conjecture two formulas
that apply to entanglement-assisted quantum convolutional codes.

\end{abstract}
\volumeyear{2008}
\volumenumber{ }
\issuenumber{ }
\eid{ }
\date{\today}
\received{\today}

\revised{}

\accepted{}

\published{}

\startpage{1}
\endpage{ }
\maketitle

The entanglement-assisted stabilizer formalism provides a useful framework for
constructing quantum codes \cite{science2006brun}. It has a number of
advantages over the standard stabilizer formalism \cite{thesis97gottesman},
including the ability to produce a quantum code from any classical linear code.

Entanglement is a valuable resource, and we would like to optimize the amount
that a sender and receiver need to consume for quantum coding. Hsieh, Devetak,
and Brun previously addressed this issue by determining a useful formula that
gives the optimal number of ebits required by a Calderbank-Shor-Steane
(CSS)\ entanglement-assisted code \cite{hsieh:062313}. In particular, the
number $c$\ of ebits that a CSS\ entanglement-assisted code requires is%
\begin{equation}
c=\mathrm{rank}\left(  HH^{T}\right)  , \label{eq:CSS-form}%
\end{equation}
where $H$ corresponds to the parity check matrix of a classical binary block
code that we import to correct quantum bit and phase flips. The same authors
also determined an upper bound for the number of ebits that an
entanglement-assisted quantum LDPC\ code requires \cite{aqis2007hsieh}.

In this Brief Report, we present several generalizations of the above formula.
Our first theorem gives a formula for the optimal number of ebits that an
arbitrary (non-CSS)\ entanglement-assisted quantum block code requires. This
formula is also a bipartite entanglement measure for a stabilizer state,
because it is equivalent to the measure in Ref.~\cite{arx2004fattal}. We find
special cases of this formula that apply to an entanglement-assisted quantum
block code produced from two arbitrary classical binary block codes, and to
codes from a classical block code over $GF(4)$. Our last formula applies to
\textit{continuous-variable} entanglement-assisted codes \cite{pra2007wildeEA}%
. We finally conjecture two formulas for the optimal number of ebits that a
general (non-CSS) entanglement-assisted quantum convolutional code or one
imported from a classical quaternary convolutional code
\cite{arx2007wildeEAQCC}\ need to consume per frame of operation. We show that
this conjecture holds for a particular example.

\begin{theorem}
Suppose we want to build an entanglement-assisted quantum code from generators
corresponding to the rows in a quantum check matrix%
\begin{equation}
H=\left[  \left.
\begin{array}
[c]{c}%
H_{Z}%
\end{array}
\right\vert
\begin{array}
[c]{c}%
H_{X}%
\end{array}
\right]  ,
\end{equation}
where $H$ is an $\left(  n-k\right)  \times2n$-dimensional binary matrix
representing the quantum code \cite{thesis97gottesman}, and both $H_{Z}$ and
$H_{X}$ are $\left(  n-k\right)  \times n$-dimensional binary matrices. Then
the resulting code is an $\left[  \left[  n,k+c;c\right]  \right]  $
entanglement-assisted code and requires $c$ ebits, where%
\begin{equation}
c=\mathrm{rank}\left(  H_{X}H_{Z}^{T}+H_{Z}H_{X}^{T}\right)  /2,
\label{eq:optimal-ebits}%
\end{equation}
and addition is binary.
\end{theorem}

\begin{proof}
Consider the matrix $\Omega_{H}=H_{X}H_{Z}^{T}+H_{Z}H_{X}^{T}$. Its entries
are the symplectic products between all rows of $H$ so that%
\begin{equation}
\left[  \Omega_{H}\right]  _{ij}=h_{i}\odot h_{j},
\end{equation}
where $h_{i}$ is the $i^{\text{th}}$ row of $H$ and $\odot$ denotes the
symplectic product \cite{arx2006brun}. Both $h_{i}$ and $h_{j}$ are each
$2n$-dimensional binary vectors and we write them as follows:%
\begin{align}
h_{i}  &  =\left(  h_{i}^{Z}\ |\ h_{i}^{X}\right)  ,\\
h_{j}  &  =\left(  h_{j}^{Z}\ |\ h_{j}^{X}\right)  ,
\end{align}
so that the first $n$ components are the \textquotedblleft Z\textquotedblright%
\ part and the last $n$ components are the \textquotedblleft
X\textquotedblright\ part. Then the symplectic product $h_{i}\odot h_{j}$ is
equal to%
\begin{equation}
h_{i}\odot h_{j}=h_{i}^{Z}\cdot h_{j}^{X}+h_{i}^{X}\cdot h_{j}^{Z},
\end{equation}
where $\cdot$ is the standard inner product and addition is binary. The matrix
$\Omega_{H}$ is a $\left(  n-k\right)  \times\left(  n-k\right)  $-dimensional
binary matrix. We call $\Omega_{H}$ the \textquotedblleft symplectic product
matrix\textquotedblright\ for the purposes of this Brief Report.

Refs.~\cite{arx2006brun,science2006brun}\ outline a symplectic Gram-Schmidt
orthogonalization procedure (SGSOP)\ that uniquely determines the optimal
(i.e., minimal) number of ebits that the code requires, and
Ref.~\cite{unpub2007got} proves that the SGSOP gives the optimal number of
ebits. The code construction in Refs.~\cite{arx2006brun,science2006brun} shows
that the resulting entanglement-assisted quantum code requires at most $c$
ebits. The essence of the argument in Ref.~\cite{unpub2007got} is that the
resulting entanglement-assisted quantum code requires at least $c$ ebits
because any fewer ebits would not be able to resolve the anticommutativity of
the generators on Alice's side of the code.

The SGSOP performs row operations that do not change the error-correcting
properties of the quantum code (because the code is additive), but these row
operations do change the symplectic product relations. These row operations
are either a row swap $S\left(  i,j\right)  $, where $S\left(  i,j\right)  $
is a full-rank $\left(  n-k\right)  \times\left(  n-k\right)  $ matrix that
swaps row $i$ with $j$, or a row addition $A\left(  i,j\right)  $, where
$A\left(  i,j\right)  $ is a full-rank $\left(  n-k\right)  \times\left(
n-k\right)  $ matrix that adds row $i$ to row $j$. These row operations
multiply the matrix $H$ from the left. The SGSOP then is equivalent to a
full-rank $\left(  n-k\right)  \times\left(  n-k\right)  $ matrix $G$ that
contains all of the row operations and produces a new quantum check matrix
$H^{\prime}=GH$ with corresponding symplectic product matrix $\Omega
_{H^{\prime}}=G\left(  H_{X}H_{Z}^{T}+H_{Z}H_{X}^{T}\right)  G^{T}$. In
particular, the resulting symplectic product matrix $\Omega_{H^{\prime}}$\ is
in a standard form so that%
\begin{equation}
\Omega_{H^{\prime}}=%
{\displaystyle\bigoplus\limits_{i=1}^{c}}
J\oplus%
{\displaystyle\bigoplus\limits_{j=1}^{n-k-2c}}
\left[  0\right]  ,
\end{equation}
where the small and large $\oplus$ correspond to the direct sum operation, $J$
is the matrix%
\begin{equation}
J=%
\begin{bmatrix}
0 & 1\\
1 & 0
\end{bmatrix}
,
\end{equation}
and $\left[  0\right]  $ is the one-element zero matrix. Each matrix $J$ in
the direct sum corresponds to a \textit{symplectic pair} and has rank two.
Each symplectic pair corresponds to exactly one ebit that the code requires
\cite{arx2006brun,science2006brun}. Each matrix $\left[  0\right]  $ has rank
zero and corresponds to an ancilla qubit. The optimal number of ebits required
for the code is $\mathrm{rank}\left(  \Omega_{H^{\prime}}\right)  /2$:%
\begin{align}
\text{rank}\left(  \Omega_{H^{\prime}}\right)   &  =\text{rank}\left(
{\displaystyle\bigoplus\limits_{i=1}^{c}}
J\oplus%
{\displaystyle\bigoplus\limits_{j=1}^{n-k-2c}}
\left[  0\right]  \right) \\
&  =\sum_{i=1}^{c}\text{rank}\left(  J\right)  +\sum_{j=1}^{n-k-2c}%
\text{rank}\left(  \left[  0\right]  \right) \\
&  =2c.
\end{align}
The second line follows because the rank of a direct sum is the sum of the
individual matrix ranks, and the third line follows from the individual matrix
ranks given above. The number $c$\ of ebits is also equal to $\mathrm{rank}%
\left(  \Omega_{H}\right)  /2$ because the matrix $G$ is full rank. The code
is an $\left[  \left[  n,k+c;c\right]  \right]  $ entanglement-assisted
quantum block code by the construction in
Refs.~\cite{arx2006brun,science2006brun}.
\end{proof}

Our formula (\ref{eq:optimal-ebits}) is equivalent to the formula at the top
of page 14 in Ref.~\cite{arx2006brun}, but it provides the quantum code
designer with a quick method to determine how many ebits an
entanglement-assisted code requires, by simply \textquotedblleft plugging
in\textquotedblright\ the generators of the code.

The formula (\ref{eq:optimal-ebits}), like the CSS\ formula in
(\ref{eq:CSS-form}), is a measure of how far a set of generators is from being
a commuting set, or equivalently, how far it is from giving a standard
stabilizer code.

Corollary~\ref{cor:CSS} below gives a formula for the optimal number of ebits
required by a CSS\ entanglement-assisted quantum code. It is generally a bit
less difficult to compute than the above formula in (\ref{eq:optimal-ebits}).
This reduction in complexity occurs because of the special form of a
CSS\ quantum code and because the size of the matrices involved are generally
smaller for a CSS\ code than for a general code with the same number of
generators and physical qubits.

\begin{corollary}
\label{cor:CSS}Suppose we import two classical $\left[  n,k_{1},d_{1}\right]
$ and $\left[  n,k_{2},d_{2}\right]  $ binary codes with respective parity
check matrices $H_{1}$ and $H_{2}$ to build an entanglement-assisted quantum
code. The resulting code is an $\left[  \left[  n,k_{1}+k_{2}-n+c,\min\left(
d_{1},d_{2}\right)  ;c\right]  \right]  $ entanglement-assisted code, and
requires $c$ ebits where%
\begin{equation}
c=\mathrm{rank}\left(  H_{1}H_{2}^{T}\right)  .
\end{equation}

\end{corollary}

\begin{proof}
The quantum check matrix has the following form:%
\begin{equation}
H=\left[  \left.
\begin{array}
[c]{c}%
H_{1}\\
0
\end{array}
\right\vert
\begin{array}
[c]{c}%
0\\
H_{2}%
\end{array}
\right]  .
\end{equation}
The symplectic product matrix $\Omega_{H}$ is then%
\begin{align}
\Omega_{H}  &  =%
\begin{bmatrix}
H_{1}\\
0
\end{bmatrix}%
\begin{bmatrix}
0 & H_{2}^{T}%
\end{bmatrix}
+%
\begin{bmatrix}
0\\
H_{2}%
\end{bmatrix}%
\begin{bmatrix}
H_{1}^{T} & 0
\end{bmatrix}
\\
&  =%
\begin{bmatrix}
0 & H_{1}H_{2}^{T}\\
H_{2}H_{1}^{T} & 0
\end{bmatrix}
.
\end{align}
The above matrix is equivalent by a full rank permutation matrix to the matrix
$H_{1}H_{2}^{T}\oplus H_{2}H_{1}^{T}$, so the rank of $\Omega_{H}$ is%
\begin{align}
\text{rank}\left(  \Omega_{H}\right)   &  =\text{rank}\left(  H_{1}H_{2}%
^{T}\oplus H_{2}H_{1}^{T}\right) \\
&  =\text{rank}\left(  H_{1}H_{2}^{T}\right)  +\text{rank}\left(  H_{2}%
H_{1}^{T}\right) \\
&  =2~\text{rank}\left(  H_{1}H_{2}^{T}\right)
\end{align}
The second line follows because the rank of a direct sum is equivalent to the
sum of the individual ranks, and the third line follows because the rank is
invariant under matrix transposition. The number of ebits required for the
resulting entanglement-assisted quantum code is rank$\left(  H_{1}H_{2}%
^{T}\right)  $, using the result of the previous theorem. The construction in
Refs.~\cite{arx2006brun,science2006brun} produces an $\left[  \left[
n,k_{1}+k_{2}-n+c,\min\left(  d_{1},d_{2}\right)  ;c\right]  \right]  $
entanglement-assisted quantum block code.
\end{proof}

\begin{corollary}
Suppose we import an $\left[  n,k,d\right]  _{4}$\ classical code over
$GF\left(  4\right)  $ with parity check matrix $H$ for use as an
entanglement-assisted quantum code according to the construction in
Ref.~\cite{arx2006brun,science2006brun}. Then the resulting quantum code is an
$\left[  \left[  n,2k-n+c;c\right]  \right]  $ entanglement-assisted quantum
code where $c=rank\left(  HH^{\dag}\right)  $ and $\dag$ denotes the conjugate
transpose operation over matrices in $GF\left(  4\right)  $.
\end{corollary}

\begin{proof}
Ref.~\cite{science2006brun} shows how to produce an entanglement-assisted
quantum code from the parity check matrix $H$\ of a classical code over
$GF\left(  4\right)  $. The resulting quantum parity check matrix $H_{Q}$\ in
symplectic binary form is%
\begin{equation}
H_{Q}=\gamma\left(
\begin{bmatrix}
\omega H\\
\bar{\omega}H
\end{bmatrix}
\right)  ,
\end{equation}
where $\gamma$ denotes the isomorphism between elements of $GF\left(
4\right)  $ and symplectic binary vectors that represent Pauli matrices.
Specifically, $\gamma^{-1}\left(  h\right)  =\omega h_{x}+\bar{\omega}h_{z}$
where $h$ is a symplectic binary vector and $h_{x}$ and $h_{z}$ denote its
\textquotedblleft X\textquotedblright\ and \textquotedblleft
Z\textquotedblright\ parts respectively. The symplectic product between binary
vectors is equivalent to the trace product of their $GF\left(  4\right)  $
representations (see, e.g., Ref.~\cite{science2006brun}):%
\begin{equation}
h_{i}\odot h_{j}=\text{tr}\left\{  \gamma^{-1}\left(  h_{i}\right)
\cdot\overline{\gamma^{-1}\left(  h_{j}\right)  }\right\}  ,
\end{equation}
where $h_{i}$ and $h_{j}$ are any two rows of $H_{Q}$, $\cdot$ denotes the
inner product, the overbar denotes the conjugate operation, and tr$\left\{
x\right\}  =x+\bar{x}$ denotes the trace operation over elements of $GF\left(
4\right)  $. We exploit these correspondences to write the symplectic product
matrix $\Omega_{H_{Q}}$ for the quantum check matrix $H_{Q}$ as follows:%
\begin{align}
\Omega_{H_{Q}}  &  =\text{tr}\left\{
\begin{bmatrix}
\omega H\\
\bar{\omega}H
\end{bmatrix}%
\begin{bmatrix}
\omega H\\
\bar{\omega}H
\end{bmatrix}
^{\dag}\right\} \\
&  =\text{tr}\left\{
\begin{bmatrix}
\omega H\\
\bar{\omega}H
\end{bmatrix}%
\begin{bmatrix}
\bar{\omega}H^{\dag} & \omega H^{\dag}%
\end{bmatrix}
\right\} \\
&  =\text{tr}\left\{
\begin{bmatrix}
HH^{\dag} & \bar{\omega}HH^{\dag}\\
\omega HH^{\dag} & HH^{\dag}%
\end{bmatrix}
\right\} \\
&  =\text{tr}\left\{
\begin{bmatrix}
1 & \bar{\omega}\\
\omega & 1
\end{bmatrix}
\otimes HH^{\dag}\right\} \\
&  =%
\begin{bmatrix}
1 & \bar{\omega}\\
\omega & 1
\end{bmatrix}
\otimes HH^{\dag}+%
\begin{bmatrix}
1 & \omega\\
\bar{\omega} & 1
\end{bmatrix}
\otimes\bar{H}H^{T}%
\end{align}
where the \textquotedblleft tr\textquotedblright\ operation above is an
element-wise trace operation over $GF\left(  4\right)  $ (it is not the matrix
trace operation.) The matrix $\Omega_{H_{Q}}$ is over the field $GF\left(
2\right)  $, but we can consider it as being over the field $GF\left(
4\right)  $ without changing its rank. Therefore, we can multiply it by
matrices over the field $GF\left(  4\right)  $. Consider the following
full-rank $GF\left(  4\right)  $\ matrices:%
\begin{equation}
A_{1}=%
\begin{bmatrix}
1 & \bar{\omega}\\
0 & 1
\end{bmatrix}
\otimes I,\ \ \ \ A_{2}=%
\begin{bmatrix}
1 & 0\\
1 & 1
\end{bmatrix}
\otimes I.
\end{equation}
We premultiply and postmultiply the matrix $\Omega_{H_{Q}}$ as follows and
obtain a matrix with the same rank as $\Omega_{H_{Q}}$:%
\begin{align}
A_{2}A_{1}\Omega_{H_{Q}}A_{1}^{\dag}A_{2}^{\dag}  &  =%
\begin{bmatrix}
0 & 0\\
0 & 1
\end{bmatrix}
\otimes HH^{\dag}+%
\begin{bmatrix}
1 & 0\\
0 & 0
\end{bmatrix}
\otimes\bar{H}H^{T}\\
&  =%
\begin{bmatrix}
\bar{H}H^{T} & 0\\
0 & HH^{\dag}%
\end{bmatrix}
\\
&  =\bar{H}H^{T}\oplus HH^{\dag}%
\end{align}
Therefore, the rank of $\Omega_{H_{Q}}$ is%
\begin{align}
\text{rank}\left(  \Omega_{H_{Q}}\right)   &  =\text{rank}\left(  \bar{H}%
H^{T}\oplus HH^{\dag}\right) \\
&  =\text{rank}\left(  \bar{H}H^{T}\right)  +\text{rank}\left(  HH^{\dag
}\right) \\
&  =2\text{ rank}\left(  HH^{\dag}\right)  .
\end{align}
The second line holds because the rank of a direct sum is the sum of the
individual ranks and the third holds because the rank is invariant under the
matrix transpose operation. Therefore, the resulting entanglement-assisted
quantum code requires $c=~$rank$\left(  HH^{\dag}\right)  $ ebits by applying
the result of the original theorem. The construction in
Refs.~\cite{arx2006brun,science2006brun}\ produces an $\left[  \left[
n,2k-n+c;c\right]  \right]  $ entanglement-assisted quantum code.
\end{proof}

\begin{corollary}
We can construct a continuous-variable entanglement-assisted quantum code from
generators corresponding to the rows in quantum check matrix $H=\left[
\left.
\begin{array}
[c]{c}%
H_{Z}%
\end{array}
\right\vert
\begin{array}
[c]{c}%
H_{X}%
\end{array}
\right]  $ where $H$ is $\left(  n-k\right)  \times2n$-dimensional, $H$\ is a
real matrix representing the quantum code \cite{pra2007wildeEA}, and both
$H_{Z}$ and $H_{X}$ are $\left(  n-k\right)  \times n$-dimensional. The
resulting code is an $\left[  \left[  n,k+c;c\right]  \right]  $
continuous-variable entanglement-assisted code and requires $c$ entangled
modes where%
\begin{equation}
c=\mathrm{rank}\left(  H_{X}H_{Z}^{T}-H_{Z}H_{X}^{T}\right)  /2.
\end{equation}

\end{corollary}

\begin{proof}
The proof is similar to the proof of the first theorem but requires
manipulations of real vectors instead of binary vectors. See
Ref.~\cite{pra2007wildeEA} for details of the symplectic geometry required for
continuous-variable entanglement-assisted codes.
\end{proof}

\begin{remark}
A similar formula holds for entanglement-assisted qudit codes by replacing the
subtraction operation above with subtraction modulo $d$. Specifically, we can
construct a qudit entanglement-assisted quantum code from generators
corresponding to the rows in check matrix\ $H=\left[  \left.
\begin{array}
[c]{c}%
H_{Z}%
\end{array}
\right\vert
\begin{array}
[c]{c}%
H_{X}%
\end{array}
\right]  $ whose matrix entries are elements of the finite field
$\mathbb{Z}_{d}$. The code requires $c$ edits (a $d$-dimensional state
$\left(  \sum_{i=0}^{d-1}\left\vert i\right\rangle \left\vert i\right\rangle
\right)  /\sqrt{d}$) where%
\[
c=\mathrm{rank}\left(  H_{X}H_{Z}^{T}\ominus_{d}H_{Z}H_{X}^{T}\right)  /2
\]
and $\ominus_{d}$ is subtraction modulo $d$. We use subtraction modulo $d$
because the symplectic form over $d$-dimensional variables includes
subtraction modulo $d$.
\end{remark}

We can also consider the case when the binary, $d$-variable, or real matrix
$H$ specifies one party's generators of a respective bipartite qubit, qudit,
or continuous-variable\ stabilizer state. In that case, our formula is
equivalent to the entanglement measure found in Ref.~\cite{arx2004fattal}.

We finally conjecture two formulas for the number of ebits required in a
general (non-CSS) entanglement-assisted quantum convolutional code.

\begin{conjecture}
The optimal number\ $c$ of ebits necessary per frame for an
entanglement-assisted quantum convolutional code is%
\begin{equation}
c=\mathrm{rank}\left(  H_{X}(D)H_{Z}^{T}(D^{-1})+H_{Z}(D)H_{X}^{T}%
(D^{-1})\right)  /2
\end{equation}
where $H(D)=\left[  \left.
\begin{array}
[c]{c}%
H_{Z}(D)
\end{array}
\right\vert
\begin{array}
[c]{c}%
H_{X}(D)
\end{array}
\right]  $ represents the parity check matrix for a set of quantum
convolutional generators that do not necessarily form a commuting set.
\end{conjecture}

\begin{conjecture}
The optimal number $c$ of ebits required per frame for an
entanglement-assisted quantum convolutional code imported from a classical
quaternary convolutional code with parity check matrix $H\left(  D\right)  $
is%
\begin{equation}
c=\mathrm{rank}\left(  H(D)H^{\dag}(D^{-1})\right)  .
\end{equation}

\end{conjecture}

We know the number of ebits required for a CSS entanglement-assisted quantum
convolutional code is$\ \mathrm{rank}\left(  H_{1}(D)H_{2}^{T}(D^{-1})\right)
$ \cite{arx2007wildeEAQCC}\ where $H_{1}(D)$ and $H_{2}(D)$ correspond to the
classical binary convolutional codes that we import to correct respective bit
and phase flips. Comparing the CSS\ convolutional formula, the CSS\ block
formula in\ Corollary~\ref{cor:CSS}, and the general block formula in
Theorem~1, the above conjecture seems natural.

We finally provide an example of the above conjecture. It is a slight
modification of the code presented in Ref.~\cite{arx2008wilde}.

\begin{example}
Consider the quantum convolutional code with quantum check matrix as follows:%
\[
\left[  \left.
\begin{array}
[c]{ccccc}%
0 & 0 & 0 & 0 & 0\\
h\left(  D\right)  & D & 0 & 1 & h\left(  D\right) \\
0 & 0 & D & D & D\\
0 & \frac{1}{D} & 1 & \frac{1}{D} & 0\\
0 & \frac{1}{D} & 0 & 0 & 0
\end{array}
\right\vert
\begin{array}
[c]{ccccc}%
h\left(  D\right)  & 0 & D & 1 & h\left(  D\right) \\
0 & 0 & 0 & 0 & 0\\
0 & 1 & 0 & 1 & 1\\
0 & 0 & 0 & 0 & 0\\
0 & 0 & 1 & 0 & 0
\end{array}
\right]  ,
\]
where $h\left(  D\right)  =1+D$. This code requires two ebits and one ancilla
qubit for quantum redundancy and encodes two information qubits. The shifted
symplectic product matrix $H_{X}(D)H_{Z}^{T}(D^{-1})+H_{Z}(D)H_{X}^{T}%
(D^{-1})$ \cite{arx2007wilde,arx2007wildeEAQCC}\ for this code is as follows:%
\begin{equation}%
\begin{bmatrix}
0 & 1 & 0 & 0 & 0\\
1 & 0 & 0 & 0 & 0\\
0 & 0 & 0 & 0 & 0\\
0 & 0 & 0 & 0 & 1\\
0 & 0 & 0 & 1 & 0
\end{bmatrix}
\end{equation}
The rank of the above matrix is four and the code requires two ebits.
Therefore, the conjecture holds for this example.
\end{example}

MMW and TAB acknowledge support from NSF Grants CCF-0545845 and\ CCF-0448658.

\bibliographystyle{apsrev}
\bibliography{minimum-ebit-formula}

\end{document}